\documentclass[amsmath, a4paper, prba, twocolumn, showpacs]{revtex4}

\usepackage{amssymb}
\usepackage{amsmath}
\usepackage{epsfig}
\usepackage{color}
\usepackage{graphics, graphicx}
\usepackage{psfrag}
\usepackage{mathcomp}
\usepackage{bbold}
\usepackage{dsfont}
\usepackage{subfigure}

\begin{document}

\author{Michiel Snoek}
\affiliation{Institute for Theoretical
Physics, University of Amsterdam, 1090 GL Amsterdam, The Netherlands}
\date{\today}

\title{Collective modes of a strongly interacting Bose gas: probing the Mott transition}

\begin{abstract}
We analyze the collective modes of a harmonically trapped, strongly interacting Bose gas in an optical lattice in the vicinity of the Mott-insulator transition. For that aim we employ the dynamical Gutzwiller equations, by performing real-time evolution and by solving the equations in linear response. We find a strong dependence on the spatial dimension of the system:
 while in one dimension the frequency of the dipole mode vanishes at the Mott transition, in higher dimensions the dominant dipole mode is featureless and we  find a signature only in the breathing mode. We discuss implications for experiments with bosonic and fermionic atoms.
\end{abstract}

\maketitle

\section{Introduction}
Both bosonic and fermionic particles in a lattice undergo the Mott-insulator transition at integer filling, when the repulsion between the particles exceeds a critical value. 
There is ample experimental evidence for the existence of those strongly correlated states of matter. 
The bosonic Mott insulator transition has been observed in systems of ultracold atoms, trapped in optical lattices \cite{Greiner02}. The fermionic Mott state is well known in solid state systems. Signatures of the fermionic Mott transition in an optical lattice have also been reported recently \cite{Jordens08, Schneider08}.

\begin{figure}
\includegraphics[width=8cm]{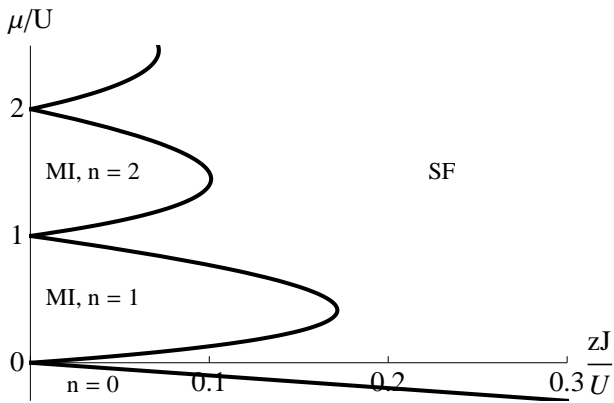}
\caption{Zero-temperature phase diagram of the Bose-Hubbard model resulting from the Gutzwiller mean-field approximation. 
The phase diagram includes a superfluid (SF) phase and Mott insulating phases (MI).
}
\label{fig_pd}
\end{figure}

The zero-temperature phase diagram of the Bose-Hubbard model is well-known \cite{Fisher89} and reproduced in Fig.~\ref{fig_pd}.
It consists of superfluid phases when the ratio of the on-site repulsion to the hopping constant is small, and Mott insulating phases when this ratio is sufficiently large. Within the Mott phases, the on-site particle number is constant and integer. 
At non-zero temperature the superfluid has a phase transition --- and the Mott phases a cross-over --- to the normal phase.
 
There are two observables by which this phase diagram is characterized: compressibility and off-diagonal superfluid long-range order. The high temperature normal phase is compressible and non-superfluid. The low-temperature superfluid phase possesses off-diagonal long-range order, and is also compressible. The zero temperature Mott phase is incompressible and not superfluid. 
There are thus in principle two possible phase transitions: one involving the existence of superfluid long-range order and one involving the compressibility.
However, these two transitions coincide at the zero temperature bosonic Mott transition, where both observables change: superfluid long-range order disappears and the system becomes incompressible.

To confirm this experimentally, it is desirable to have independent probes for both of these transitions. This need is even stronger for the case of fermionic atoms, in which case only the compressible-incompressible transition takes place at the Mott transition.

In past experiments with bosonic atoms, primarily the superfluid-normal transition has been confirmed. This follows directly from the investigation of interference peaks after time-of-flight. 
The compressible-incompressible transition in bosonic atoms has been probed by the observation of the wedding-cake structure \cite{Folling06, Campbell06, Gemelke09}: for values of the bosonic repulsion exceeding the critical interaction, and sufficiently high filling, a Mott plateau appears surrounded by a superfluid \cite{Jaksch98}.  For even higher filling a second superfluid emerges in the center, which reaches a new Mott plateau at still higher filling, thus giving rise to a series of alternating superfluids with smootly varying density and Mott plateaus with constant integer density.

For fermionic atoms, two other methods have been used to probe the reduction of the compressibility at the Mott insulator transition: the observation of the cloud size \cite{Schneider08} and the fraction of double occupied sites \cite{Jordens08}.
Both methods, however, have the disadvantage that in an inhomogeneous system, the Fermi liquid regions of the system also contribute to the signal. This implies that even at zero temperature these observables do not show a sharp feature at the Mott insulating transition. This makes it difficult to use these methods to locate the Mott transition precisely.

In this article we propose to use the frequencies of the collective modes as probes of the Mott transition, in particular to probe the compressibility. We find a strong dependence on the dimension: while in one dimension the frequency of the dipole mode vanishes at the transition, in higher dimensions this clear signature disappears.
In particular, in two and three dimensions
the dominant dipole mode is featureless at the transition. 
Therefore, more subtle signatures have to be be sought. 
They are most pronounced in the breathing mode, because exciting a breathing mode indeed compresses the system, in contrast to the dipole and quadrupole mode.
However, the signal is non-generic and depends on  the choice to probe the mode frequencies while increasing the particle number at constant interaction, or while increasing the repulsion at constant particle number. 
In the former case there is no sharp signature at the onset of the Mott insulator, because the support of the breathing mode smoothly moves to the edge when the particle number is increased. There is a sharp signal when a superfluid appears on top of the Mott insulator, because then a new mode, with very low frequency appears.
When increasing the interaction at constant particle number the signal depends on the particle number. When the particle number is sufficiently high, a sharp signal is found: when approaching the onset of the Mott transition in the trap center, the breathing mode turns into a bimodal structure; the low-frequency mode, corresponding to excitations in the trap center, vanishes at the transition and only the high-frequency mode, corresponding to excitation of the edge of the system, survives.  
For lower particle numbers, however, these two modes show an avoided crossing and the sharp signal at the transition disappears.

We also find that when a Mott plateau is present in the trap center, the breathing mode and the sub-dominant dipole- and quadrupole modes approach each other. 

We provide explicit calculations of the frequency of the modes in the case of bosonic particles, for which aim we use the time-dependent Gutzwiller (mean-field) approximation. 
From this we can build up a physical picture, which enables us to extend the result to the collective mods of inhomogeneous fermionic systems in optical lattice.
We now first present the model in Sec. \ref{mod} and the method in Sec. \ref{meth}. The results are presented in Sec. \ref{res}. We close with a discussion and conclusions in Sec. \ref{conc}.

\section{Model}
\label{mod}
For a deep optical lattice and moderate filling, both fermionic and bosonic particles can be described by a single-band Hubbard Hamiltonian in tight-binding approximation. Here we consider bosons, which are described by the Bose-Hubbard model \cite{Fisher89}:
\begin{eqnarray}
\mathcal{H} &=& - J \sum_{\langle i j \rangle} \left\{ b_i^\dagger b_j + {\rm h.c.} \right\}
\\ \nonumber &&+ \sum_i \left\{\frac{U}{2} \hat n_i (\hat n_i - 1) + \left(V(i, t)  - \mu \right) \hat n(i) \right\}.
\end{eqnarray}
Here $b_i^{(\dagger)}$ is the annihilation (creation) operator at site $i$ and we defined $\hat n_i = b_i^\dagger b_i$. Furthermore, $\mu$ is the chemical potential,
$J$ is the hopping amplitude and $U$ is the on-site repulsion. $J$ and $U$ can be expressed in terms of the atomic interparticle scattering length $a$, atomic mass $m$ and laser wavelength and intensity \cite{Bloch08}. We will use them as effective parameters.  $V(i, t)$ is the underlying harmonic potential which we take equal to:
\begin{equation}
V(i ,t ) = V_0 (t) |{\bf x}_i - {\bf x}_0(t)|^2.
\end{equation}
Here we have indicated a possible time-dependence of the trap center and the trap constant, which is used to induce the collective modes.
In the following we set the lattice constant equal to $a=1$. We express the mode frequencies in terms of the frequency associated with the harmonic trap $\omega_{\rm trap} = 2\sqrt{V_0/J}$.

\section{Method}
\label{meth}

We use the time-dependent Gutzwiller mean-field approximation \cite{Jaksch02}, in which the hopping between the lattice sites is treated in a mean-field approximation. 
For inhomogeneous systems, this procedure has to be carried out in a space-resolved version, where with each site a different order parameter is associated.
The total many-body wave function is within this approximation given as
$
| \Psi \rangle = \prod_i \sum_{n=0}^{n_c} f_n^i \frac{( b_i^\dagger)^n } {\sqrt{n!}} |0 \rangle.
$ 
Here we have introduced a cut-off $n_c$ on the number of Fock states we take into account. The value of this cut-off should be chosen such that the results are independent of it, which depends on the strength of the interaction and the local density. In the strongly interacting, low filling regime we focus on here, $n_c$ could be chosen to be a small number.
The dynamics is governed by the set of coupled differential equations \cite{Jaksch02}
\begin{eqnarray}
i \dot f_n^i &=& - J \sum_{ \langle i j  \rangle }  \left( \sqrt{n+1} \; \Phi_j^* \; f_{n+1}^i + \sqrt{n} \; \Phi_j \; f_{n-1}^i \right) \nonumber \\ 
&& +  \left( \frac{U}{2} n (n-1) + V(i,t) - \mu \right) f_n^i,
\label{gutzdyn}
\end{eqnarray}
where $\Phi_i = \langle b_i \rangle = \sum_n \sqrt{n} (f_{n-1}^i)^* f_n^i$.

\begin{figure}[b]
\includegraphics[width=8cm]{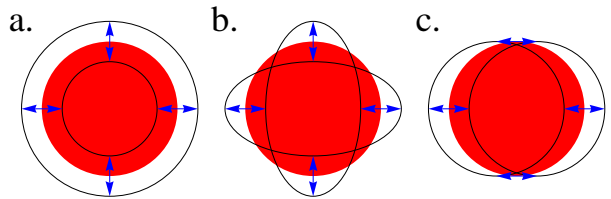}
%\hspace{-2cm}
\caption{(Color online) Symmetry of the collective modes. Here we depict the two-dimensional situation: a. breathing mode; b. quadrupole mode; c. dipole mode}
\label{figmodes}
\end{figure}

\subsection{Linear response}
When the deviation from equilibrium is small, the dynamical Gutzwiller equation can also be solved in linear response \cite{Menotti08}. After obtaining the equilibrium ground state, characterized by the coefficients $f_{n0}^i$, we write $f_n^i(t) = f_{n0}^i + \delta f_n^i(t)$. Expanding the equations of motion in Eq. (\ref{gutzdyn}) up to linear order in $\delta f_n^i(t)$ yield the linearized equations for the collective modes. However, one has to make sure that the excitations are orthogonal to the ground state and therefore first a projection to this orthogonal subspace is needed.

For a homogeneous system one can use momentum conservation to perform this procedure for each momentum separately. However, in an inhomogeneous system, as we study here, this is not possible. This means that a large matrix, of size $2 V n_c$ ($V$ being the number of lattice sites), has to be diagonalized. For a one-dimensional system this is possible for realistic system sizes, but for higher dimensions the real-time evaluation of Eq.   (\ref{gutzdyn}) turns out to be more efficient. 

\subsection{Validity}
The Gutzwiller approximation is a highly efficient method for studying dynamics of lattice bosons. It conserves energy and particle number with a very good accuracy. The latter, however, is only true if the sites are sequentially updated; a parallel update of all sites together is numerically unstable with respect to particle number conservation \cite{Snoek07}. The validity of the Gutzwiller approximation is further justified by the fact that for small interactions it incorporates the Gross-Pitaevskii dynamics \cite{Jaksch02}. 
For the fully connected lattice (i.e. the infinite-dimensional limit), the Gutzwiller mean-field equations are shown to be exact \cite{Snoek11}.

Although this mean-field approximation can thus only be justified in high-dimensional lattices, 
we apply the Gutzwiller mean-field dynamics here also to one- and two- dimensional systems. This is outside the strict regime of validity of the approximations, and therefore we do not expect quantitative accuracy. Within this mean-field approximation only local quantum fluctuations are taken into account, whereas long-wavelength fluctuations are very important in one dimension, because they destroy true superfluid long range order. For the trapped systems at zero temperature we consider this is less of an issue, because the harmonic potential provides a natural cut-off for the long-wavelength fluctuations.
  On a quantitative level 
 the critical interaction for the Mott transition is strongly renormalized compared to the mean-field prediction: the equilibrium  Gutzwiller mean-field approximation predicts the value of the Mott insulating transition for homogeneous a homogeneous system with $n$ particles per site to happen at $U/zJ = 2n+1 + 2 \sqrt{n(n+1)}$ ($z$ being the number of lattice neighbors).  For $n=1$ this gives reasonable agreement with Quantum Monte Carlo predictions for the three-dimensional cubic lattice \cite{Capogrosso07}, but for the square lattice \cite{Capogrosso08} and one-dimensional lattice \cite{Ejima11} quantum fluctuations significantly lower the critical repulsion and hence increase the Mott insulating lobes.

 However, we are not so much interested here in a quantitative determination of the critical interaction, but in the qualitative, lattice-geometry dependent, features accompanying this transition. We expect them to be not affected by quantum fluctuations.

\subsection{Deduction of the mode frequencies}
We induce the collective modes by starting the simulation with the system prepared slightly away from equilibrium as sketched in Fig. \ref{figmodes}.
The breathing mode is induced by a sudden change in the spring constant $V_0(t)$, which experimentally corresponds to a sudden change in trapping frequency. Experimentally, the breathing mode is also excited when the optical lattice is ramped up, because the increasing $U/J$ ratio pushes particles away from the center \cite{Wernsdorfer10}.
 The dipole mode is induced by a sudden change in the trap center ${\bf x}_0$ \cite{Burger01, Cataliotti01, Fertig05, McKay}. The quadrupole mode is obtained by starting with an anisotropic trap $V(i ,t ) = \sum_{\alpha} V_{\alpha} (t) ({\bf x}_\alpha - {\bf x}_{0\alpha}(t))^2$.

Note that in one dimension the quadrupole does not exist. There we can only induce the dipole and breathing mode. The breathing mode is unique; the dipole and quadrupole modes are degenerate in higher dimensions.

After following the time evolution, we deduce the modes by fitting the cloud size and center of mass-position to a sum of cosines, using the frequencies and amplitudes as fitting parameters. By taking a sum of cosines we are able to identify the dominant mode frequencies, but also sub-dominant modes can be identified.

Note that throughout this article, we always restrict ourselves to small initial perturbations and hence small deviations from equilibrium. To induce the dipole mode we shifted the trap minimum by only a tenth of the lattice constant; the monopole and quadrupole mode were induced by a one-percent change in the trap parameter.  
These small perturbations guarantee that  we are in the linear response regime, such that the absolute magnitude of the perturbation is not an independent quantity of physical interest.

\section{Results}
\label{res}

We present the results of the numerical simulations for the collective modes. Because we want to study the signal in the collective modes at the Mott insulating transition, we need to locate this transition in the equilibrium case. This is a subtle issue, because of the inhomogeneous system: the superfluid adjacent to the Mott insulator leads to an exponentially decaying superfluid order parameter, instead of one that is exactly vanishing.  
Therefore we accompany the  plots of the frequencies with pictures showing the evolution of the radial density profiles $n(r)$ and absolute value of the local superfluid order parameter squared $| \langle \hat b \rangle |^2(r)$, when the total particle number or the ratio $U/J$ is changed.

\subsection{One dimension}
First we present results in one spatial dimension. As argued before, in this case the Gutzwiller dynamics cannot be trusted quantitatively. However, one dimension allows for a comparison with beyond mean-field methods \cite{Pupillo03, Lundh04, Lundh04b, Rey05, Liu05, Montangero09}.
 Moreover, since the number of sites is limited, we can compare the numerical time evolution with the mode-frequencies from a linear response calculation even for realistic trapping frequencies,  

\begin{figure}
\vspace{-.2cm}
\includegraphics[width=7.2cm]{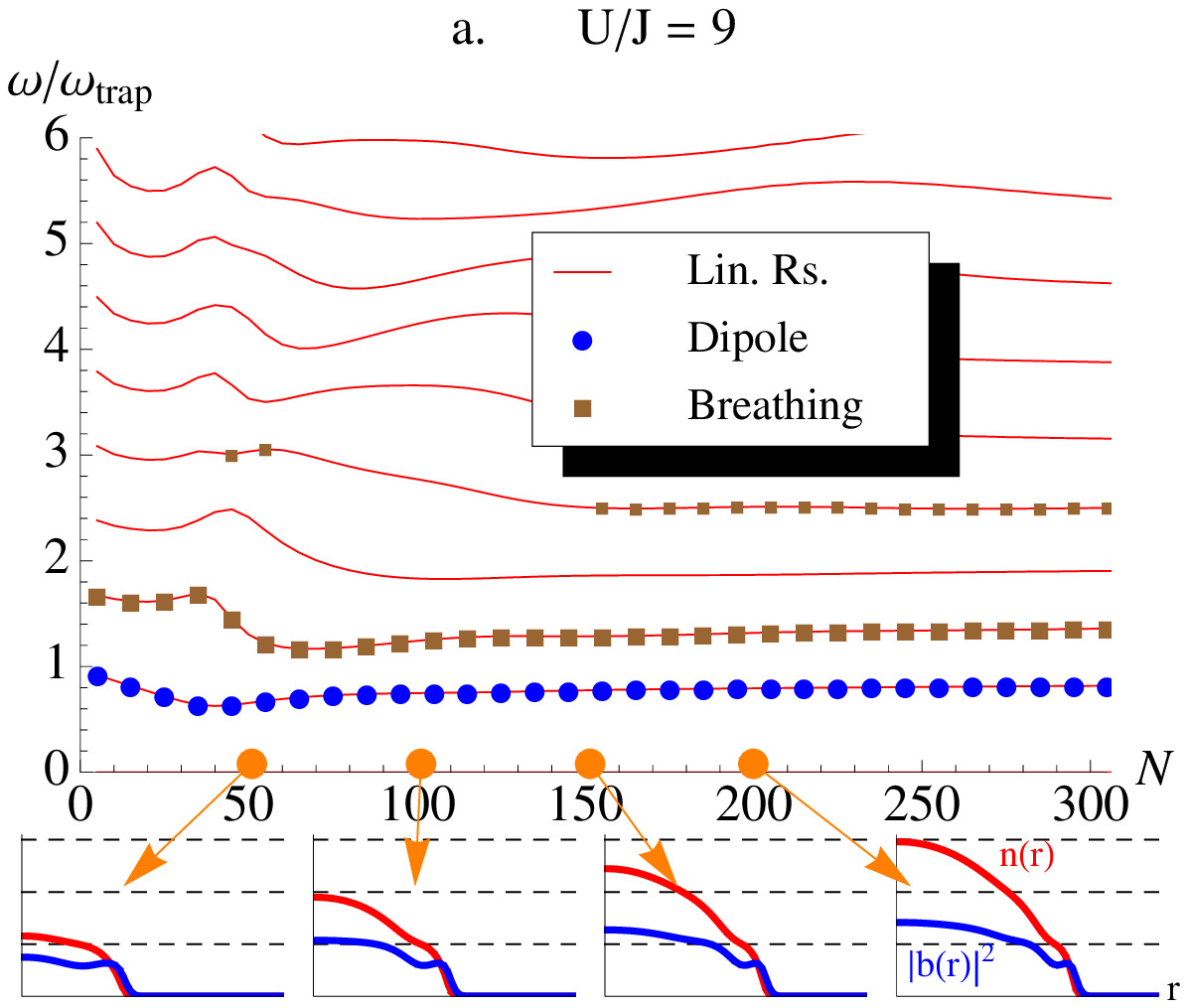}
\includegraphics[width=7.2cm]{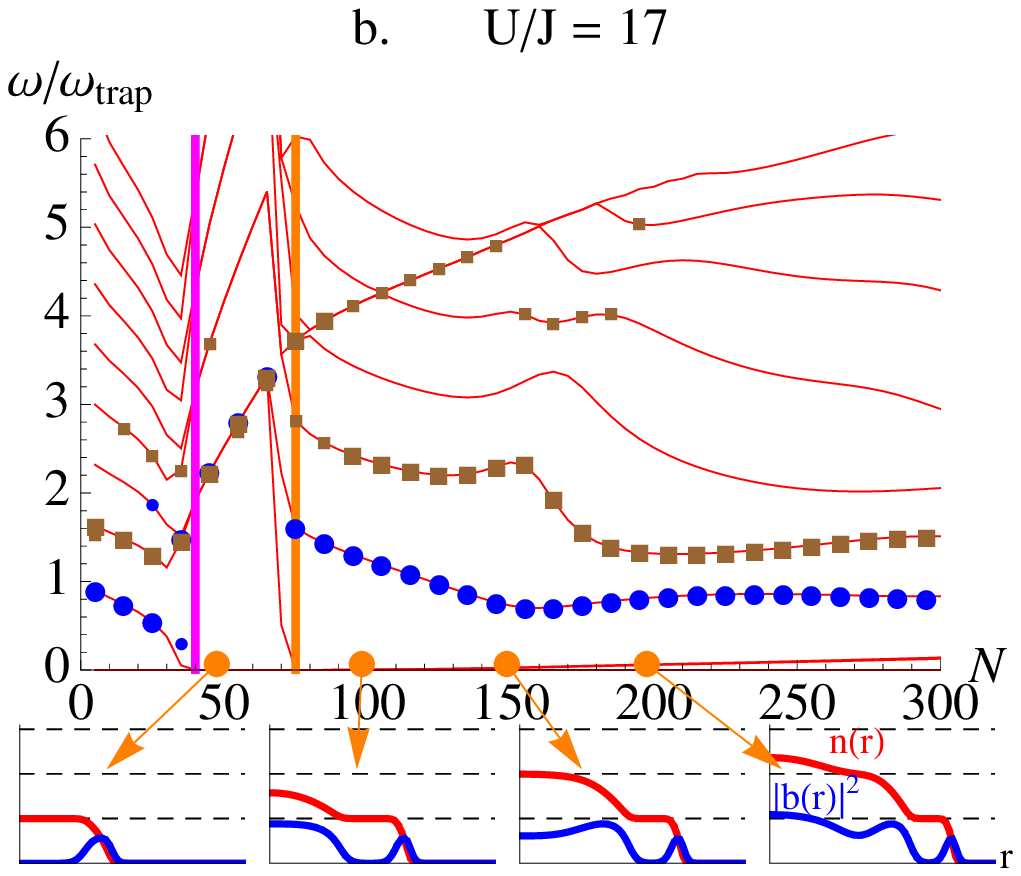}
\includegraphics[width=7.2cm]{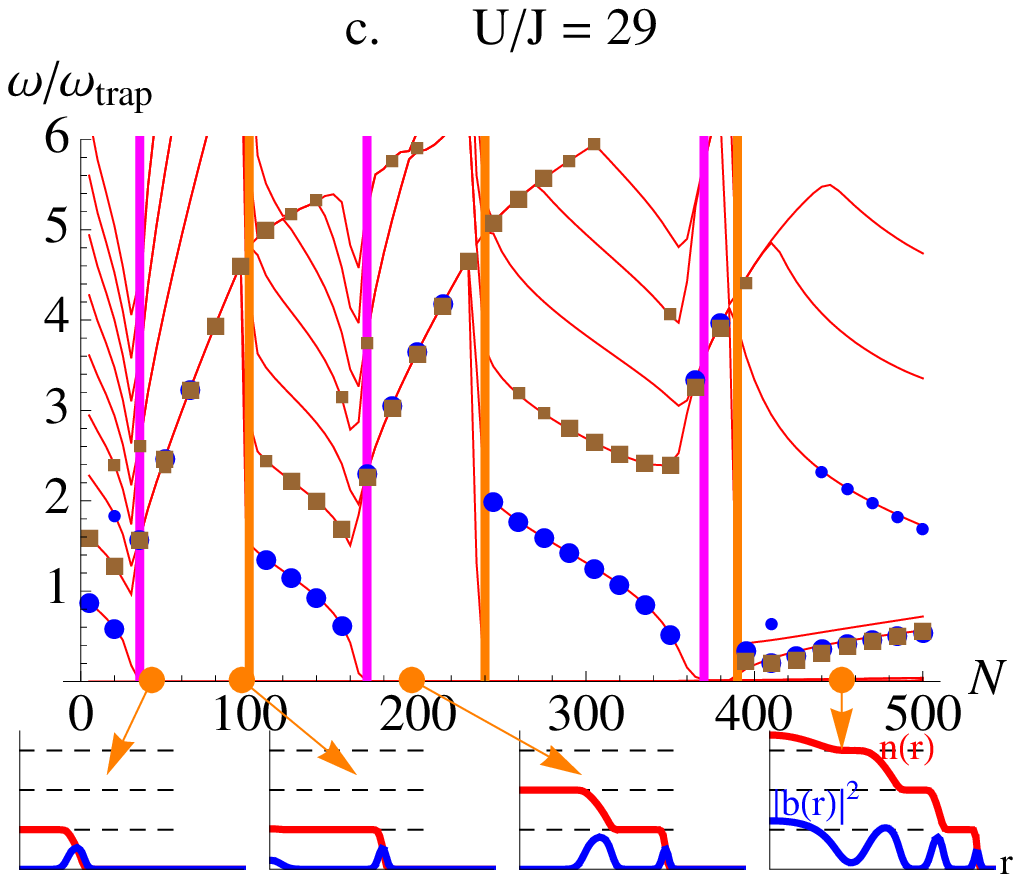}
\caption{(Color online) Frequencies of the collective modes scaled to the trapping frequency in \emph{one dimension} as a function of particle number $N$ for fixed $U/J=9, 17$ and $29$. Red lines represent the result of the linear response calculate. Blue dots and brown squares indicate the result of the numerical time-evolution after a quench inducing a dipole and breathing mode, respectively. 
The small dots/squares denote the frequencies of the subdominant modes.
The trap parameter is $V_0 = 0.01 J$. Magenta vertical lines denote the onset of a Mott insulator in the center; orange lines denote the onset of a superfluid. The small panels denote the radial profiles of the density and $| \langle \hat b \rangle |^2(r)$ for some values of $N$.
}
\label{figU}
\end{figure}

\subsubsection{Mode frequency as a function of $N$ for constant $U/J$}
We start by directing our attention to the frequency of the modes for constant $U/J$ as a function of $N$, as depicted in Fig. \ref{figU}. First of all we note the excellent agreement between the linear response calculation and the numerical time-evolution, except for a few points close to the phase transition.

To begin with, we look at the dominant frequencies, indicated by the large dots, and concentrate on the dipole mode.
For small particle numbers, i.e. in the weakly interacting limit, we observe that the dipole mode approaches the trap frequency, which coincides with the dipole frequency in the the non-interacting limit. This will be true for all further plots as well.
 Mean-field interactions indeed do not change the frequency of the dipole mode \cite{Stringari96}. However, the interactions beyond mean-field included in the dynamical Gutzwiller equations do change the frequency, as we observe for larger particle numbers, since this includes (partial) localization of the particles and depletion of the superfluid. 

In Fig. \ref{figU}a. we have chosen $U/J = 9$, which is below the critical interaction for the formation of a Mott insulator.
Still we see a minimum in the dipole mode frequency, at the point that the compressibility in the trap center is mimimal. 
When increasing $U/J$ the minimum in the dipole mode frequency becomes progressively smaller, until it completely vanishes at the onset of the Mott insulator for  $U/J > (U/J)_c$. We see this in  
Fig. \ref{figU}b. At this point, also the breathing mode frequency has a minimum.

When the particle number is increased further, the Mott plateau in the center broadens and is completely immobile. This means that the only action can happen within the superfluids at the edges, which oscillate at high frequency. We note that the dipole mode frequency and monopole frequency coincide in this region. This is because they correspond to the in-phase and out-of-phase oscillations of the two superfluids edges, which have the same frequency.

For even higher frequencies, a second superfluid emerges at the trap center. This has a dramatic effect on the frequencies: they jump to a much lower value since now the particle currents can be carried by the superfluid particles in the center.

When $U/J$ is chosen even higher, we can observe multiple Mott plateaus forming when the particle number is increased, as visible in Fig. \ref{figU}c. We see the same pattern occurring each time the Mott insulator forms in the center: the dipole mode frequency vanishes and the breathing mode frequency has a minimum, followed by high and coinciding frequencies. At the onset of the superfluid in the center, the frequencies jump down again. 

\begin{figure}
\includegraphics[width=7.2cm]{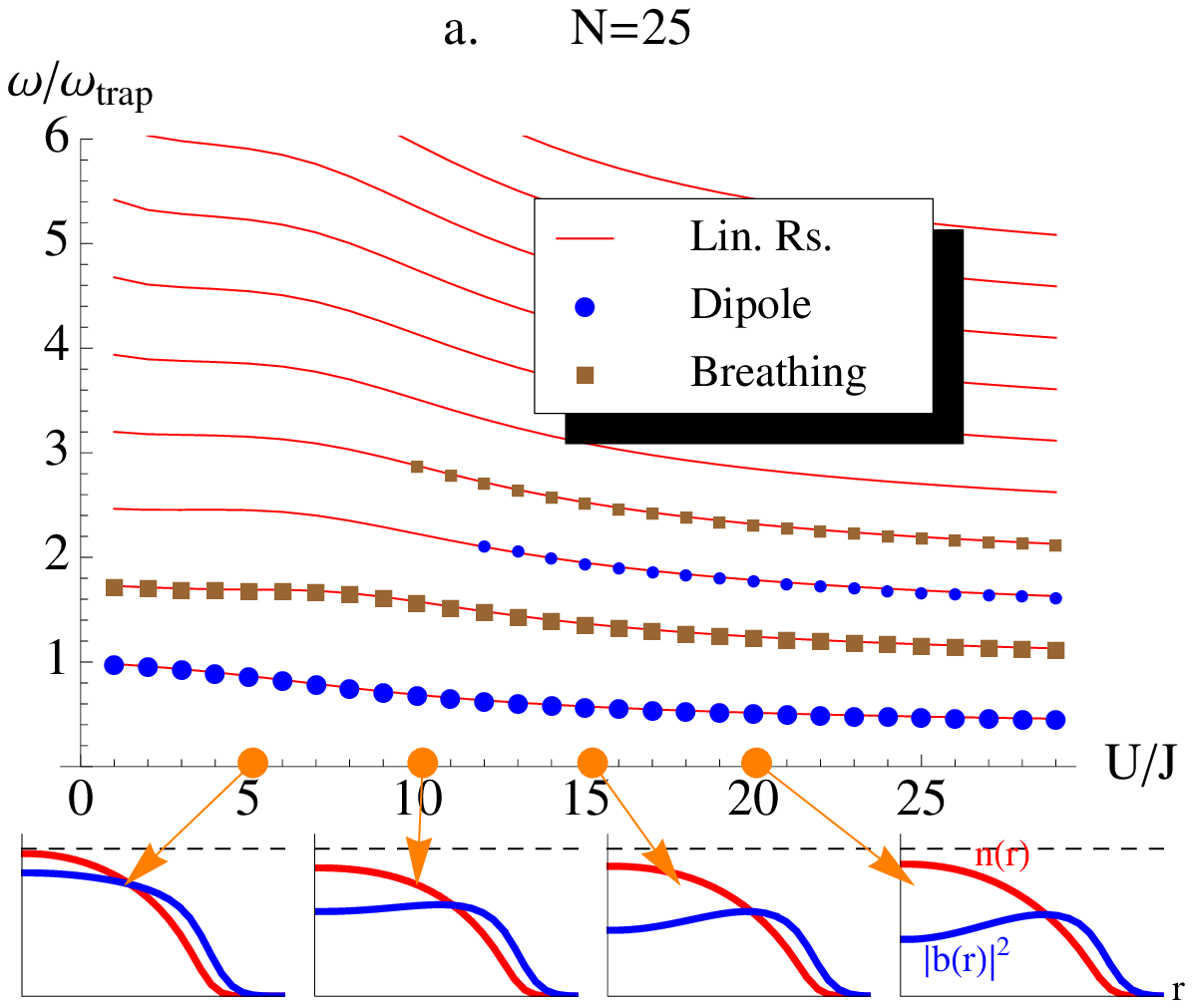}
\includegraphics[width=7.2cm]{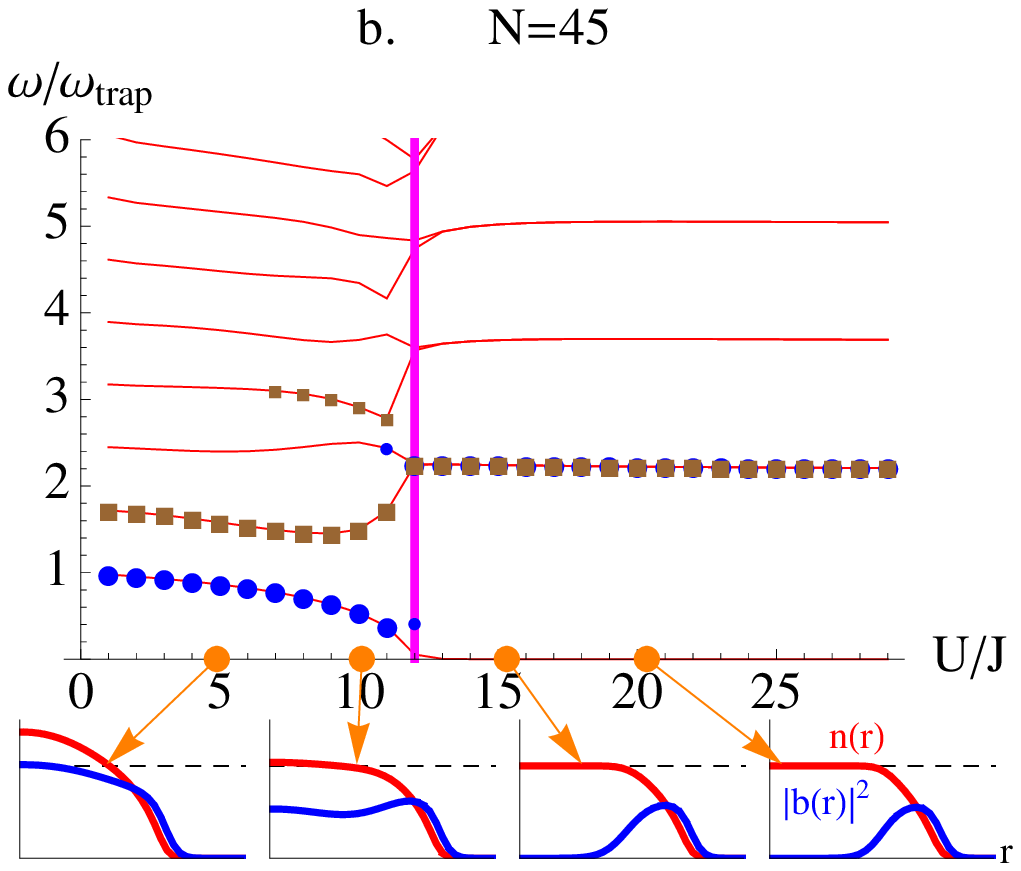}
\includegraphics[width=7.2cm]{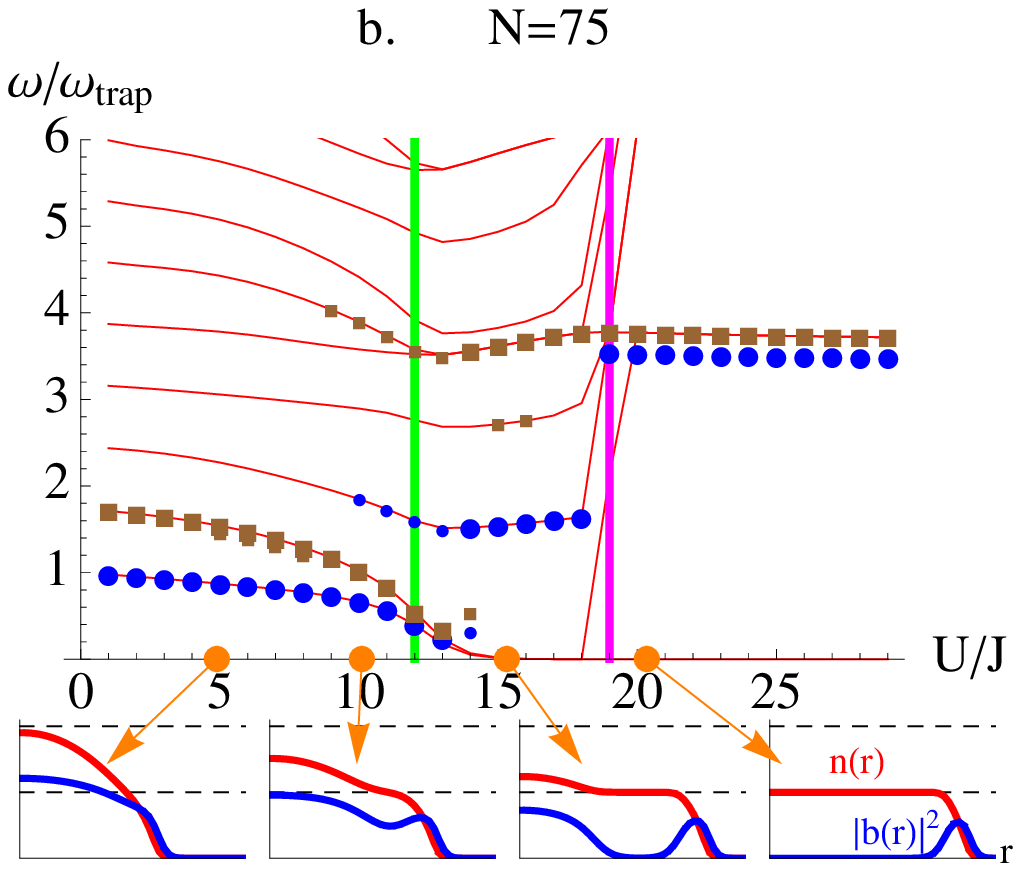}
\caption{(Color online) Frequencies of the collective modes scaled to the trapping frequency in \emph{one dimension} as a function of the interparticle repulsion $U$ for fixed particle number $N$. Red lines represent the result of the linear response calculate. Blue dots and brown squares indicate the result of the numerical time-evolution after a quench inducing a dipole and breathing mode, respectively. 
The small dots/squares denote the frequencies of the subdominant modes.
The trap parameter is $V_0 = 0.01 J$. Magenta vertical lines denote the onset of a Mott insulator in the center, the green vertical line denotes the onset of a Mott insulator away from the center. The small panels denote the radial profiles of the density and $| \langle \hat b \rangle |^2(r)$ for some values of $U/J$.
}
\label{figN}
\end{figure}

The fact that de dipole mode vanishes at the transition can be understood by the following simple argument: in the superfluid the Bogoliubov sound modes have a linear spectrum, with a sound velocity that vanishes at the onset of the Mott insulator \cite{Oosten01}. This makes that mass transport becomes increasingly slow and totally vanishes at this point. For the dipole mode, this leads to a vanishing frequency, because this mode involves particle transport across the trap center; the point where the sound velocity vanishes. However, we also observe that an increasing fraction of the particles is reflected by this almost incompressible trap center, leading to the appearance of a second frequency (indicated by smaller dots) in the dipole spectrum.
The breathing mode does not vanish at the transition, because the particles in the center, where the sound velocity vanishes, are not involved in the particle transport carrying this mode.   
The fact that the breathing mode frequencies shoot up when a Mott insulator is formed in the center, follows from the fact that then transport through the center is completely suppressed, because the modes are gapped in this region. Only the edges fluctuate with a high frequency.

We thus see that in one dimension there is a clear signal in the collective modes when the Mott transition takes place, if one scans the mode frequencies as a function of the particle number for constant $U/J$. This signal was also found for fermionic particles \cite{Liu05}. However, in that case it was observed that also the breathing mode frequency vanishes as the transition. In contrast, here we observe that although the breathing mode frequency has a minimum at the onset of the Mott insulator, it does not vanish.

\subsubsection{Mode frequency as a function of  $U/J$ for constant $N$}
We now investigate the one-dimensional mode frequencies when $N$ is kept constant, but $U/J$ is varied. Some representative plots are shown in Fig. \ref{figN}. In this case it depends strongly on the particle number what the behavior of the modes is across the Mott transition. When the particle number is too low, no Mott transition takes place, as shown in Fig. \ref{figN}a. The frequencies are smooth in this case.

When the particle number is increased, the Mott transition takes place in the center as shown in Fig. \ref{figN}b. In this case the dipole mode vanishes at the transition, but the breathing mode does not vanish, like we have seen in the previous subsection. However, we observe that the breathing mode has a kink-like structure at the transition: it sharply goes up when the transition is approached and then reaches a constant value when the Mott state is present. We can understand that the amplitude goes up before the transition, because as the transition approaches the central region (where the insulator forms) becomes more and more incompressible, thus pushing the mobile particles supporting the breathing mode to the edges, where the frequencies are naturally higher. We also observe that the mode structure at this point has the form of an avoided crossing between modes localized in the center and in at the edge.

When the particle number is even higher, the Mott transition does not take place first in the center, but in a shell surrounding the center. As a consequence, both the dipole mode and the breathing mode vanish at this transition, because in both modes particles with a vanishing sound velocity are involved. This is shown in Fig. \ref{figN}c. 

After the Mott insulator has taken place away from the center, the superfluid center and the superfluid edges are effectively decoupled. This leads to independent oscillations of the edges and the superfluid center, leading to additional low-lying modes in this region.

When the full center is Mott insulating, the mode frequencies become practically independent of $U/J$. This reflects the fact that the density profiles become independent of the repulsion for sufficiently large repulsion and sufficiently low total particle number, as also visible in the panels in Fig. \ref{figN} showing the radial density profiles. The reason for this is that also within the superfluid the fraction of doubly occupied sites at those large values of $U/J$ is very small, such that effectively the hard-core limit is reached.

We thus see that also in this scheme, the collective modes show a clear signature at the Mott insulating transition in one spatial dimension.

\subsection{Two dimensions}

\begin{figure}

\includegraphics[bb = 0 -50 300 296, width=2.8cm]{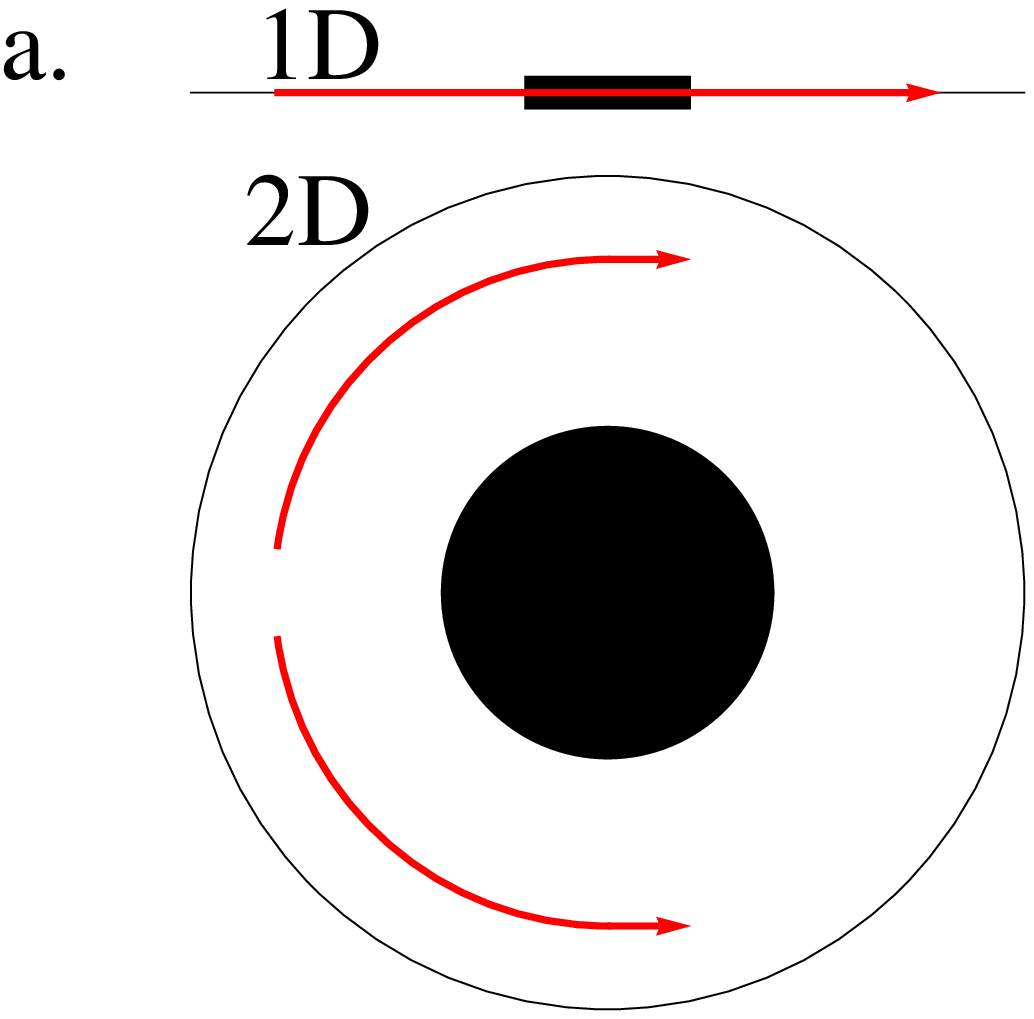}  \hspace{.2cm}
\includegraphics[width=4.2cm]{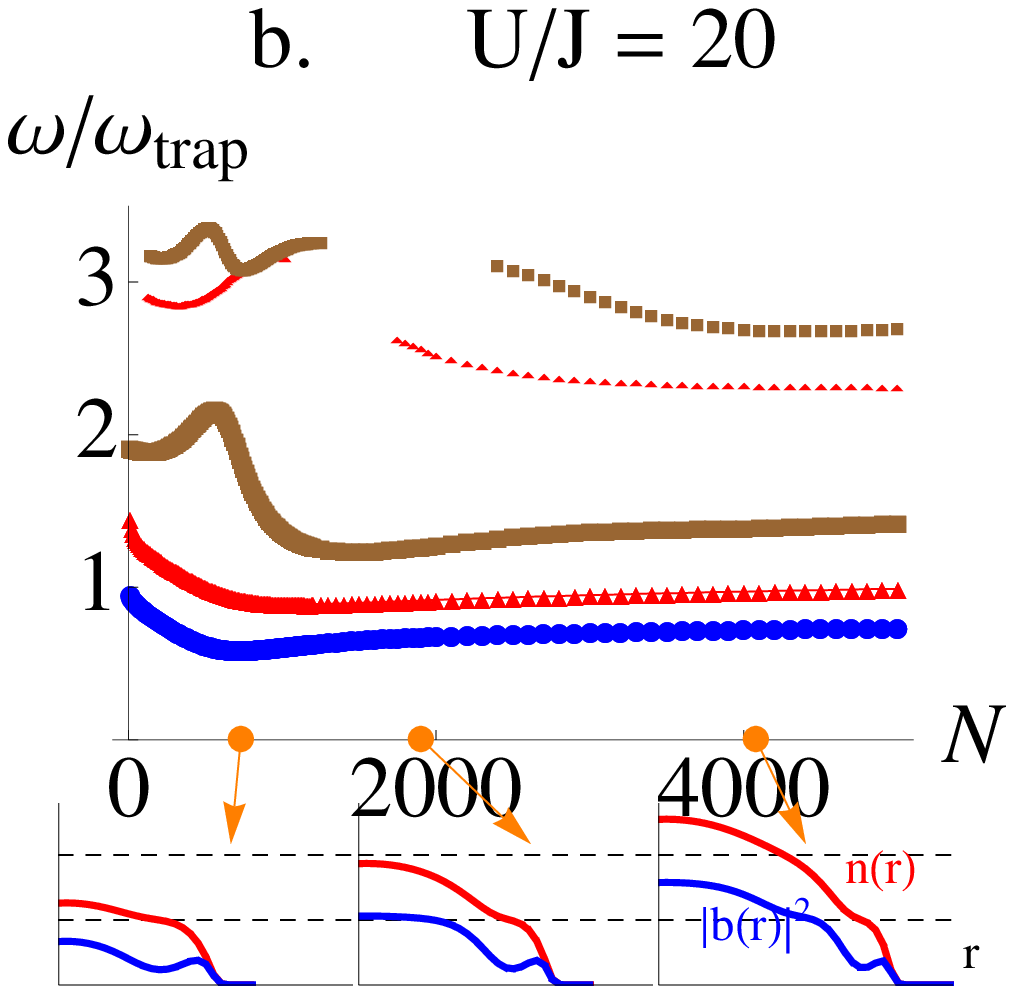} 
\vspace{.2cm}
\includegraphics[width=7.2cm]{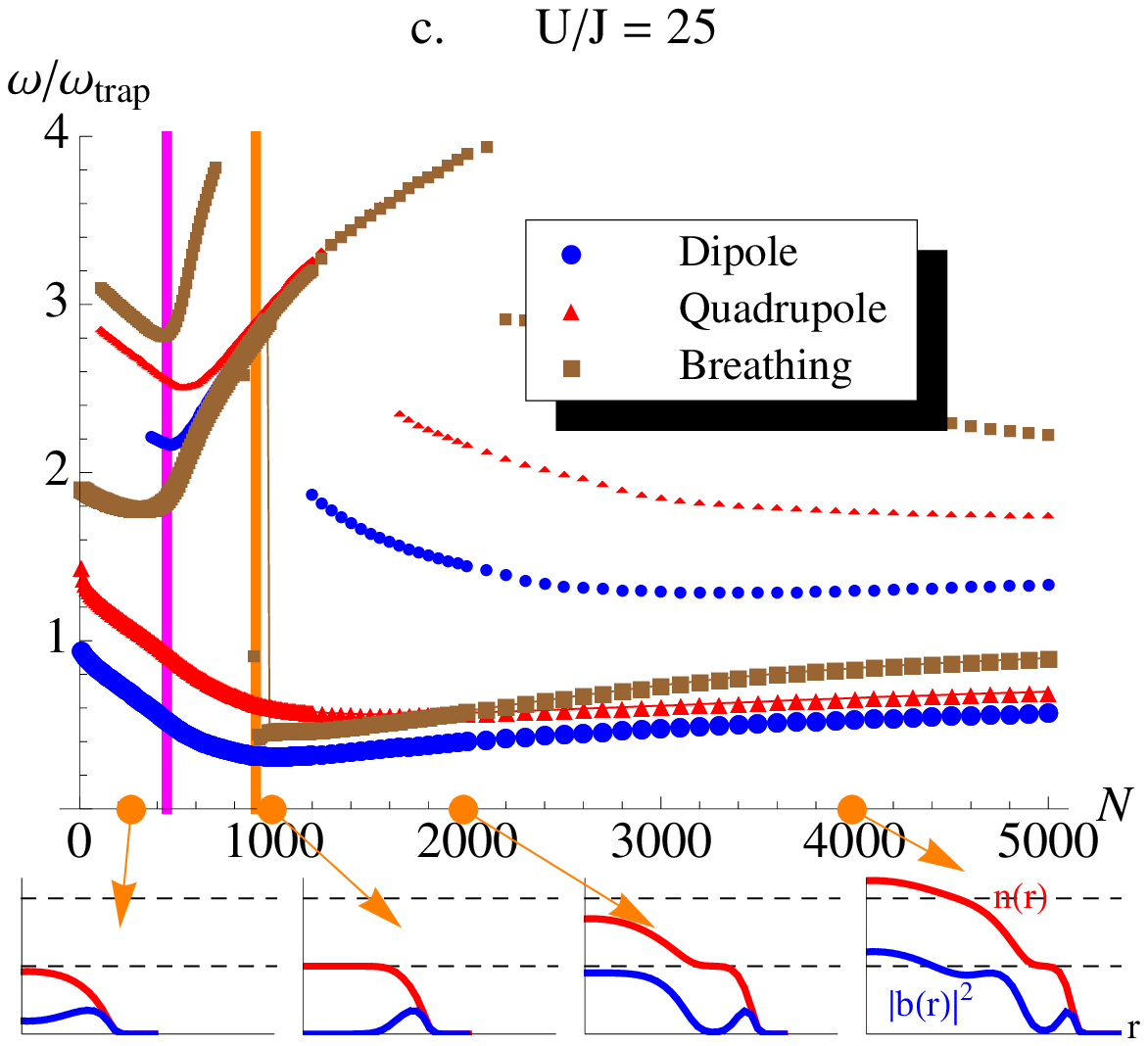} 
\includegraphics[width=7.2cm]{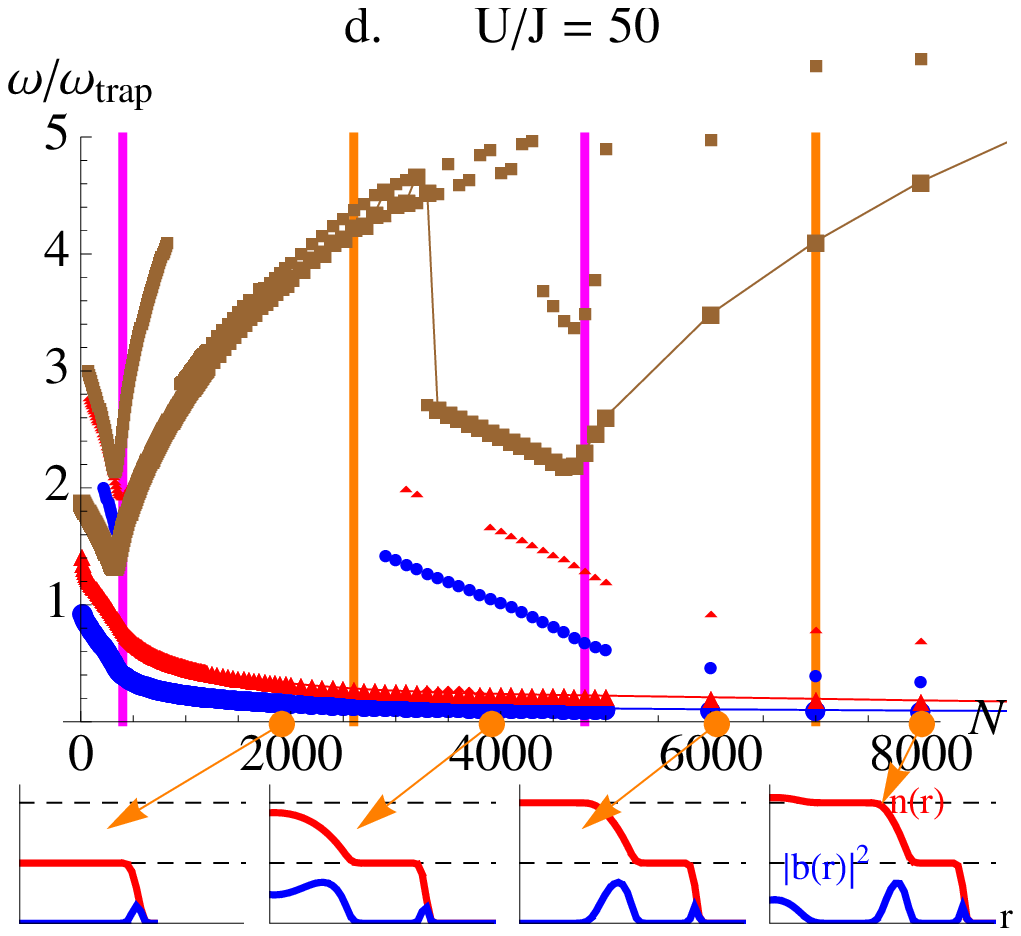}

\caption{(Color online) 
a: Difference between the dipole mode in one and two dimension: in one dimensions particles have to cross the insulating center, whereas in two (and higher) dimensions the particles can avoid it.
b-d: Frequencies of the collective modes scaled to the trapping frequency in \emph{two dimensions} as a function of particle number $N$ for fixed $U/J$. Blue dots, red triangles and brown squares indicate the dipole, quadrupole and breathing mode frequencies, respectively. 
The small dots/squares/triangles denote the frequencies of the subdominant modes.
The trap parameter is $V_0 = 0.05J$. Magenta vertical lines denote the onset of a Mott insulator in the center; orange lines denote the onset of a superfluid.
The small panels denote the radial profiles of the density and $| \langle \hat b \rangle |^2(r)$ for some values of $N$.
}
\label{fig2d_U}
\end{figure}

\begin{figure}
\includegraphics[width=7.1cm]{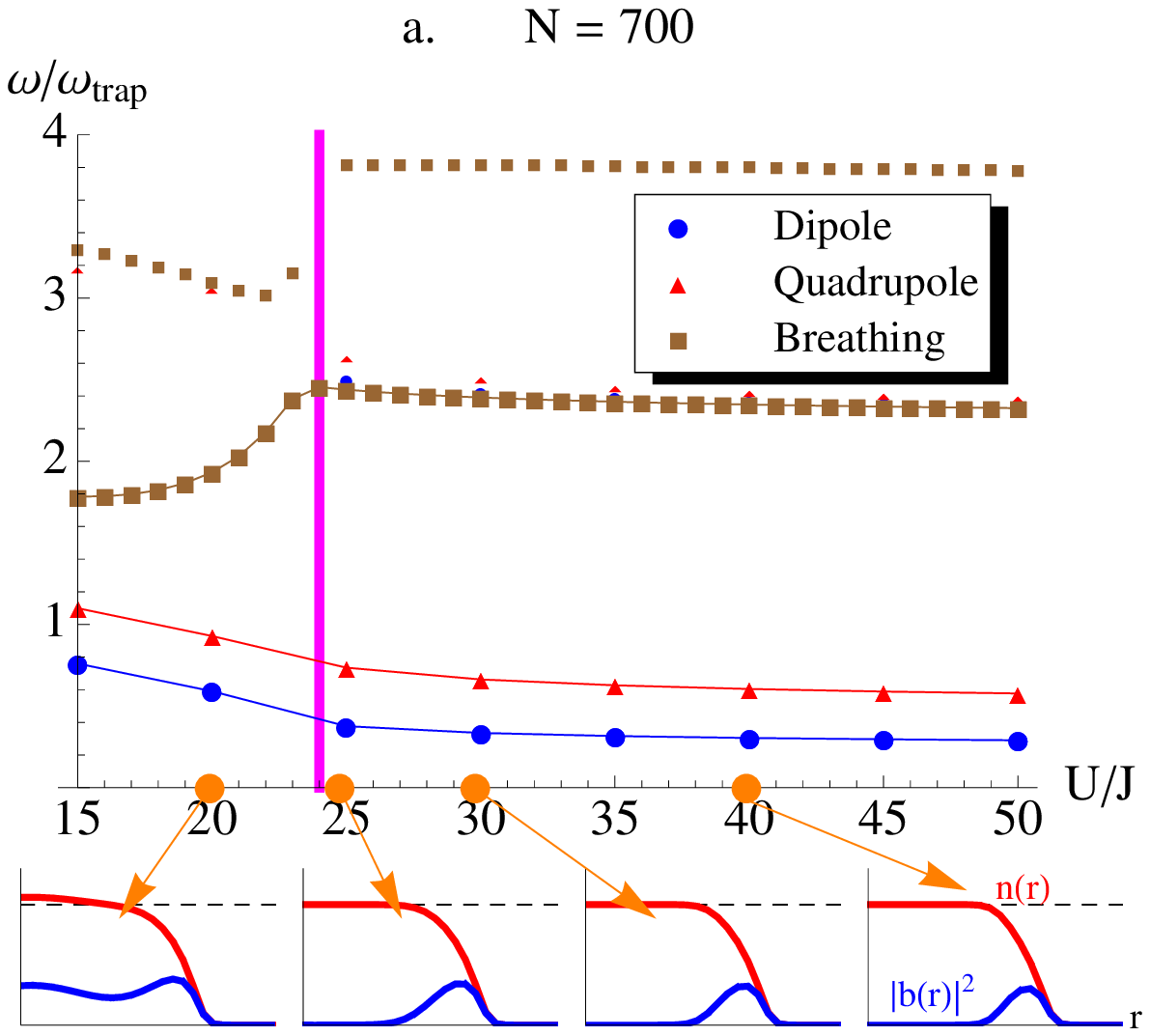} 
\includegraphics[width=7.1cm]{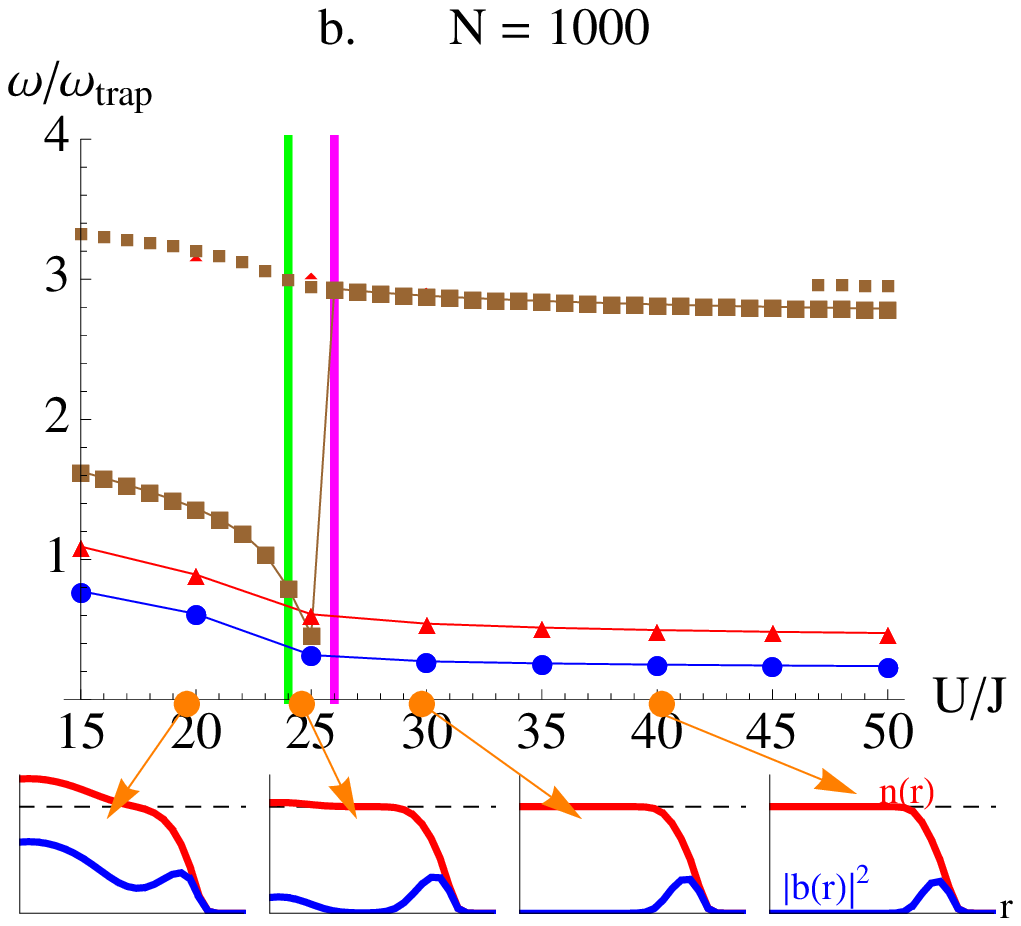} 
\includegraphics[width=7.1cm]{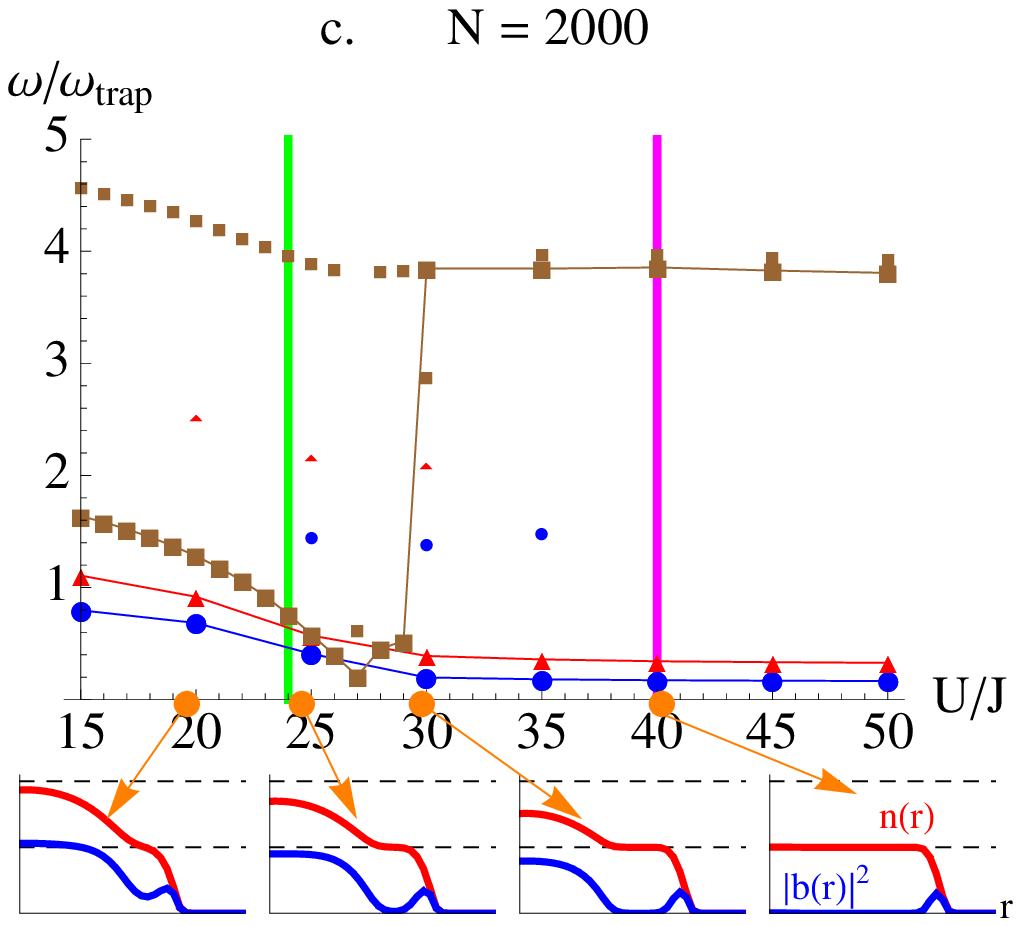} 

\caption{(Color online) Frequencies of the collective modes scaled to the trapping frequency in \emph{two dimensions} as a function of interaction $U/J$ for fixed particle number $N$. Blue dots, red triangles and brown squares indicate the dipole, quadrupole and breathing mode frequencies, respectively. 
The small dots/squares/triangles denote the frequencies of the subdominant modes.
The trap parameter is $V_0 = 0.05J$. Magenta vertical lines denote the onset of a Mott insulator in the center; green vertical lines denote the onset of a Mott insulator away from the center
The small panels denote the radial profiles of the density and $| \langle \hat b \rangle |^2(r)$ for some values of $U/J$.
}
\label{fig2d_N}
\end{figure}

The results for two dimensional systems are shown in Fig. \ref{fig2d_U} and \ref{fig2d_N}, for the case that the particle number is increased at constant $U/J$ and the particle number is kept constant but $U/J$ is increased, respectively.

We first of all note the important difference between the one-and two-dimensional case regarding the frequency of the dipole modes: whereas in one dimension the dipole mode shows a clear signal at the Mott insulator transition by completely vanishing, in two dimensions the dominant dipole mode is completely featureless at the transition. 

This is a purely geometrical effect, which can be understood in a simple way (see Fig. \ref{fig2d_U}a): in one dimension particles contributing to the dominant dipole mode have to cross the trap-center, where the Mott insulator is formed and the sound velocity tends to zero. In contrast, in two and higher dimensions, particles carrying the dominant dipole mode can circumvent the incompressible trap center. Therefore, the dominant trap frequency is not sensitive to the Mott transition at all. 
This same argument applies to the quadrupole mode.
When looking for signs of the Mott insulator transition, we therefore have to concentrate on the breathing mode. Indeed, when this mode is excited, the cloud is compressed, such that this mode should give information about the compressibility. 

\subsubsection{Mode frequencies as a function of  $N$ for constant $U/J$}
We again first investigate the case that $U/J$ is kept fixed and the particle number is increased.
This does not lead to a sharp signal in the breathing mode frequency at the onset of the Mott insulator,
as shown in Fig. \ref{fig2d_U}. Slightly before, but not exactly equal to, the Mott transition the dominant monopole mode has a minimum. After the Mott insulator has settled in the center, the dominant breathing mode and the second dipole and quadrupole mode approach each other. 

However, a very clear signal appears in this scheme when a superfluid in the center forms: this leads to the sudden appearance of low-lying collective modes in the spectrum. The nature of this low-lying mode depends on $U/J$: when the repulsion is only slightly larger than the critical repulsion (as the case in Fig. \ref{fig2d_U}c), the superfluid at the edge and in the center are strongly coupled. This leads to a mode with a very low frequency, corresponding to the coupled motion of the superfluid in the center and the edge. We observe that this mode becomes dominant immediately after the superfluid in the center forms.
In contrast, when the repulsion is chosen larger (as the case in Fig. \ref{fig2d_U}e), the superfluid at the edge and in the center are separated by a large Mott plateau and hence only weakly coupled. This means that the edge and the center in first instance oscillate independently, and the superfluid at the edge still supports the dominant mode. For larger particle number the superfluid in the center contains progressively more particles and takes over the dominant breathing mode. As a subdominant mode we then observe the appearance of a mode with a very low frequency, corresponding to the coupled in-phase motion of the superfluid in the center and at the edge. This mode becomes dominant for even larger particle numbers. For larger $U/J$ (see Fig. \ref{fig2d_U}e) we also see the Mott plateau with two localized particles per site appearing. However, because $U/J$ is so large, we do not observe the low-lying mode corresponding the coupled superfluids at the edge and in the center.

\begin{figure}
\includegraphics[width=8cm]{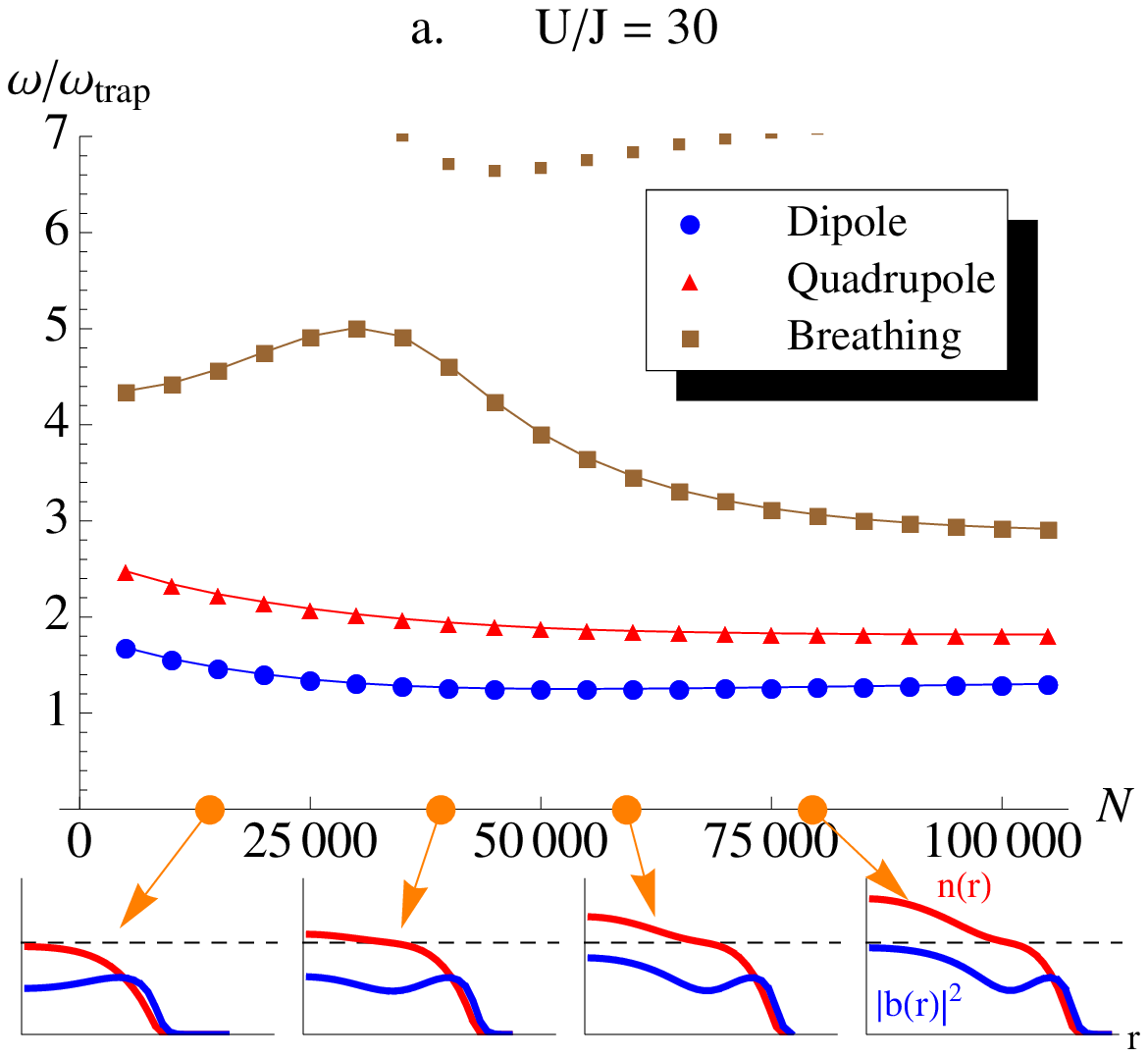} 
\includegraphics[width=8cm]{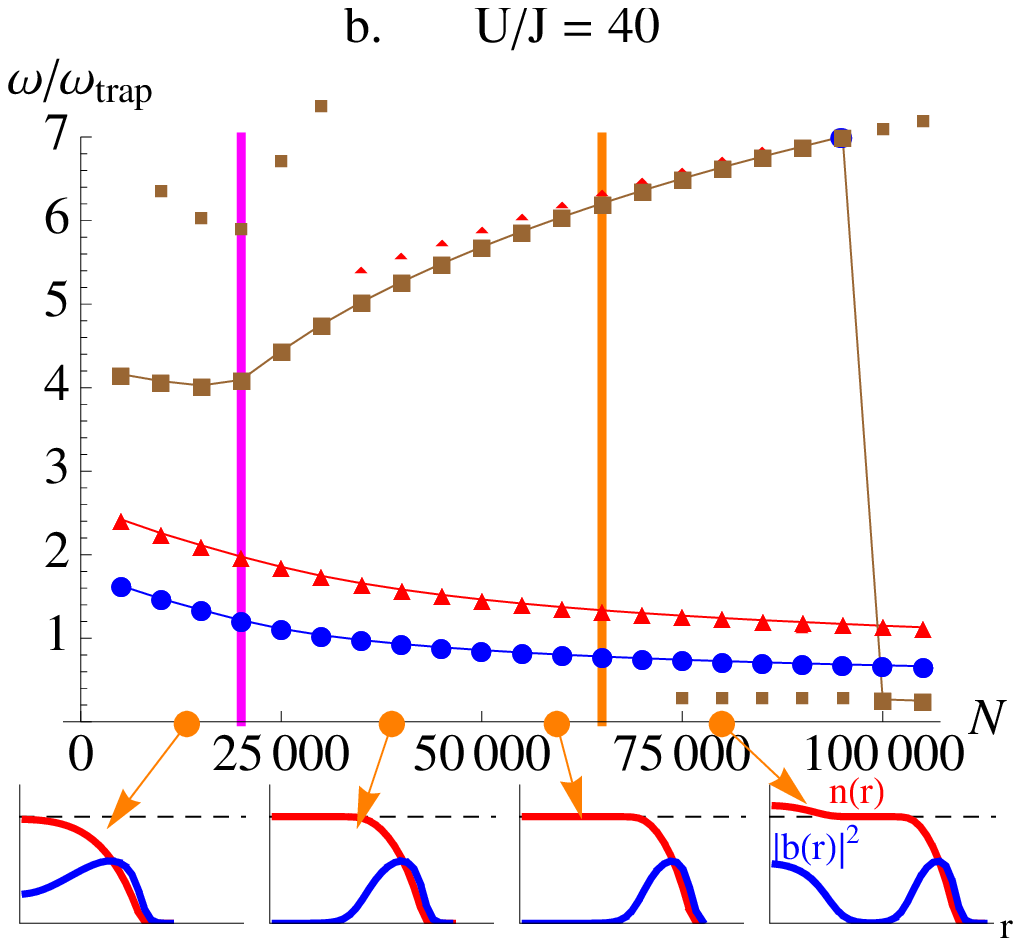} 

\caption{(Color online) Frequencies of the collective modes scaled to the trapping frequency in \emph{three dimensions} as a function of particle number $N$ for fixed $U/J$. Blue dots, red triangles and brown squares indicate the dipole, quadrupole and breathing mode frequencies, respectively. 
The small dots/squares/triangles denote the frequencies of the subdominant modes.
The trap parameter is $V_0 = 0.04J$. Magenta vertical lines denote the onset of a Mott insulator in the center; orange lines denote the onset of a superfluid.
The small panels denote the radial profiles of the density and $| \langle \hat b \rangle |^2(r)$ for some values of $N$.
}
\label{fig3d_U}
\end{figure}

\begin{figure}
\includegraphics[width=8cm]{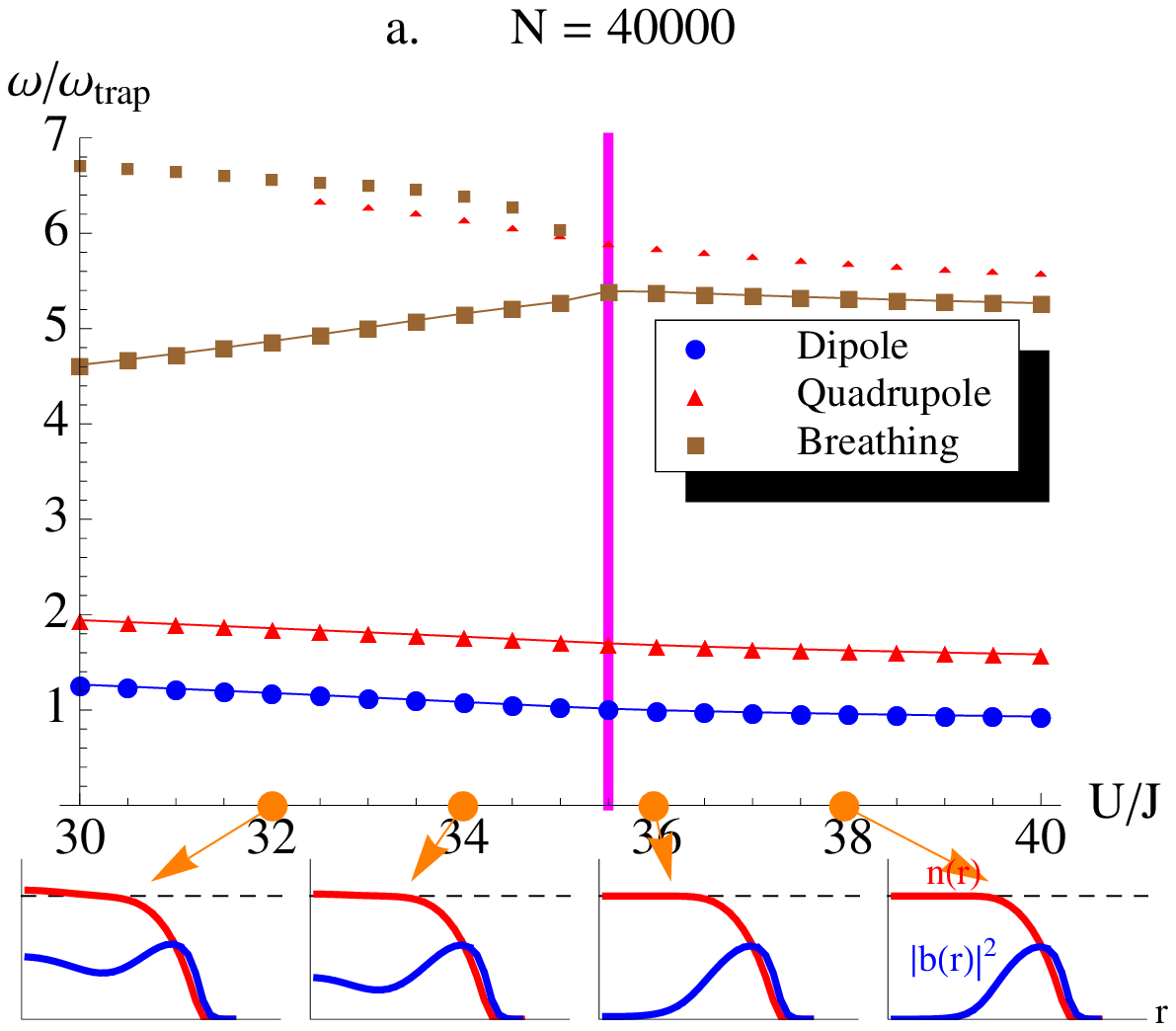} 
\includegraphics[width=8cm]{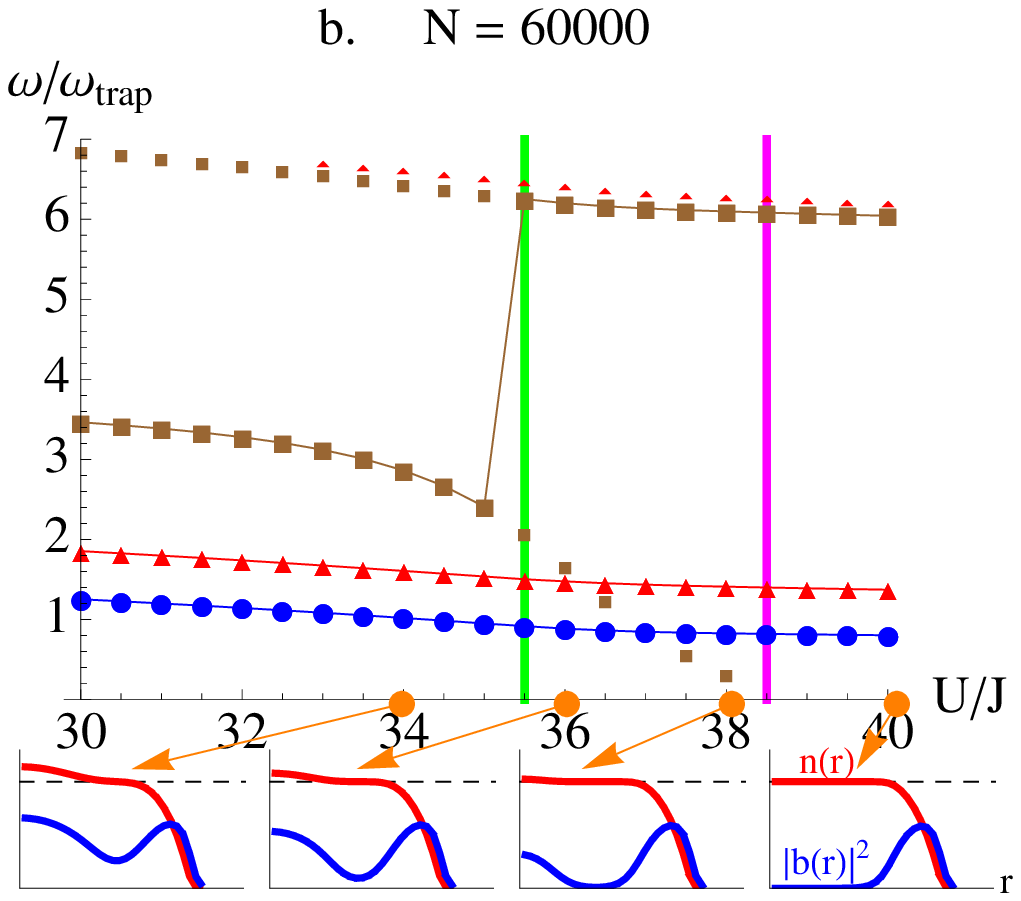} 

\caption{(Color online) Frequencies of the collective modes scaled to the trapping frequency in \emph{three dimensions} as a function of interaction $U/J$ for fixed particle number $N$. Blue dots, red triangles and brown squares indicate the  dipole, quadrupole and breathing mode frequencies, respectively. 
The small dots/squares/triangles denote the frequencies of the subdominant modes.
The trap parameter is $V_0 = 0.04J$. Magenta vertical lines denote the onset of a Mott insulator in the center; the green vertical line denotes the onset of a Mott insulator away from the center.
The small panels denote the radial profiles of the density and $| \langle \hat b \rangle |^2(r)$ for some values of $U/J$.
 }
\label{fig3d_N}
\end{figure}

 \subsubsection{Mode frequencies as a function of  $U/J$ for constant $N$}
 When the repulsion is increased for constant particle number, the two-dimensional situation is rather similar to the one-dimensional situation, as shown in Fig. \ref{fig2d_N}.
 The signal at the Mott insulator transition 
 depends on the particle number. When the particle number is sufficiently low such that the Mott transition happens in the center (see Fig. \ref{fig2d_N}a), the dominant breathing mode shows a (more-or-less sharp) kink at the transition. We observe again the avoided crossing between modes localized in the center and at the edge.
  In contrast, when the particle number is sufficiently high, such that the Mott transition happens in a shell away from the center (as the case in Fig. \ref{fig2d_N}b, c), the dominant breathing mode vanishes, thus providing a clear signal.
 However, we see that the breathing mode frequency only vanishes in this case after the Mott insulating plateau is strong enough: when the repulsion is only a little bit larger than the critical interaction, the Mott plateau is so small that particle transport through it can still happen.  

\subsection{Three dimensions}

The situation in three dimensions is similar to the two-dimensional situation and shown in the plots in Figs. \ref{fig3d_U} and \ref{fig3d_N}. This is because the topology in two- and three-dimensional is the same: only in one dimensions the two superfluids at the edge are disconnected once a Mott insulator has formed in the center and particles carrying the dipole mode have to cross the center. In all higher dimensions the superfluids form connected shells, which can sustain a dipole mode which does not have to enter the insulating regime. When the particle number is increased at constant $U/J$ we therefore see a smooth dependence of the collective modes when the repulsion is chosen smaller than the critical value for the Mott insulator transition (Fig. \ref{fig3d_U}a.), and a sudden appearance of a low frequency breathing mode at the onset of a superfluid in the center when the repulsion exceeds the critical value (Fig. \ref{fig3d_U}b.).
 Note that in Fig. \ref{fig3d_U}b. the repulsion is chosen only slightly larger than the critical interaction, such that the superfluid at the edge and in the center are coupled and a very low-lying mode appears when the superfluid appears in the center, like we have seen in Fig. \ref{fig2d_U}c. 
 
 When the repulsion is increased at constant particle number, we see a kink in the breathing mode at the onset of the Mott insulator when the particle number is chosen such that the Mott transition happens in the trap center (Fig. \ref{fig3d_N}a.). When the particle number is larger, the Mott insulator happens in a shell outside the center, leading to a sharp signal in the breathing mode: a jump of the dominant breathing mode to a high value and a subdominant breathing mode that vanishes when the Mott insulator approaches the center.

\section{Conclusions}
\label{conc}

In this article we studied the behavior of the collective modes of a trapped strongly interacting Bose gas in an optical lattice. 
Our main goal was to investigate whether this can be used as a probe for the Mott insulator transition.
We found a particularly strong signature in one spatial dimension, because there the dominant dipole mode completely vanishes at the transition. This is no longer true in two (and higher) spatial dimensions, where the dominant dipole mode is completely featureless at the transition. 

The strongest signature in two (and higher) dimensions appears in the breathing mode when the particle number is sufficiently high: when $U/J$ is increased at constant particle number the breathing mode vanishes at the transition, followed by a jump to a high value, indicating that the mode is completely supported by the edge of the system. When the repulsion is kept constant and the particle number is increased, there is no sharp signal in the collective modes at the onset of the Mott insulator. However, the appearance of a superfluid on top of the insulator leads to the sudden appearance of low-lying modes and thus a clear signal.

Although our simulations are performed for bosonic particles, we can use these results to predict the behavior of fermionic particles at the Mott transition as well. Namely, since we can understand the behavior of the modes in terms of the compressibility, we expect similar behavior for fermionic atoms. Also for fermonic atoms we therefore predict that the dipole mode vanishes completely in one dimension at the Mott Insulator  transition (as indeed confirmed \cite{Liu05}), but is featureless in higher dimensions. We expect signatures in the breathing mode in the same fashion as found for bosons: a vanishing breathing mode frequency at sufficiently high particle number when the repulsion is increased through the insulator transition and a sharp down-jump of the breathing mode frequency at the point that a metallic region appears on top of the Mott insulator when the particle number is increased at constant repulsion. 

\section*{Acknowledgements}

This 
work was supported by the Nederlandse Organisatie voor Wetenschappelijk Onderzoek (NWO).

\end{document}